 \documentclass[preprint2]{aastex}
\usepackage{graphics}
\usepackage{graphicx}
\usepackage{multirow}
\usepackage{amssymb}
\usepackage{color}
\usepackage{boxedminipage}

\def \kms {{\rm km/s}}

\newcommand{\hmpc}{{\,\rm h^{-1}Mpc}}

\shorttitle{Cosmic Flows-1}
\shortauthors{Courtois et al.}

\begin{document}

\title{3D Velocity and Density Reconstructions of the Local Universe with Cosmicflows-1}

\author{H\'el\`ene M. Courtois$^{1,2}$}
\affil{$^1$Institute for Astronomy (IFA), University of Hawaii, 2680 Woodlawn Drive, HI 96822, USA}
\affil{$^2$University of Lyon; UCB Lyon 1/CNRS/IN2P3/INSU; IPN Lyon, France} 
\email{courtois@ifa.hawaii.edu}
\author{Yehuda Hoffman$^3$}
\affil{$^2$Racah Institute of Physics, Hebrew University, Jerusalem 91904, Israel} 
\author{R. Brent Tully$^1$}
\affil{$^1$Institute for Astronomy (IFA), University of Hawaii, 2680 Woodlawn Drive, HI 96822, USA }
\author{Stefan Gottl\"ober$^4$}
\affil{$^4$Leibniz-Institut f\"ur Astrophysik Potsdam (AIP), An der Sternwarte 16, 14482 Potsdam, Germany}

\begin{abstract}
This paper presents an analysis of the local peculiar velocity field based on the Wiener Filter reconstruction method. 
We used our currently available catalog of distance measurements containing 1,797 galaxies within 3000 km/s: Cosmicflows-1.  The Wiener Filter method is used to recover the full 3D peculiar velocity field from the observed map of radial velocities and to recover the underlying linear density field.  The velocity field within a data zone of 3000 km/s is decomposed into a local component that is generated within the data zone and a tidal one that is generated by the mass distribution outside that zone.  The tidal component is characterized by a coherent flow toward the Norma-Hydra-Centaurus (Great Attractor) region while the local component is dominated by a flow toward the Virgo Cluster and away from the Local Void.  A detailed analysis shows that the local flow is predominantly governed by the Local Void and the Virgo Cluster plays a lesser role.   
The analysis procedure was tested against a mock catalog.  It is demonstrated that the Wiener Filter accurately recovers the input velocity field of the mock catalog on the scale of the extraction of distances and reasonably recovers the velocity field on significantly larger scales.   The Bayesian Wiener Filter reconstruction is carried out within the $\Lambda$CDM WMAP5 framework.

The Wiener Filter reconstruction draws particular attention to the importance of voids in proximity to our neighborhood.  The prominent structure of the Local Supercluster is wrapped in a horseshoe collar of under density with the Local Void as a major component.

\end{abstract}


\footnote{Accepted for publication in ApJ , August 6th,  2011}

\section{Introduction}
\label{sec:intro}

This paper presents a reconstruction of the large scale structure of the nearby Universe from a survey of galaxy peculiar velocities using the means of the Wiener Filter (WF) and constrained realizations (CR). The observational component is a catalog of distances and inferred line-of-sight peculiar velocities: Cosmicflows-1, data assembled by \citet{2008ApJ...676..184T}.   This information can be used to derive three-dimensional velocity and density fluctuation fields with use of the Wiener filter \citep{1995ApJ...449..446Z, 2009LNP...665..565H}.  The subsequent description of phase space in the local region can be used to infer initial conditions that can be the starting point for constrained numerical simulations \citep{2010arXiv1005.2687G}.

The goal of the program is to strengthen the link between an improving knowledge of the observed positions and motions of nearby galaxies and the increasingly sophisticated theoretical renditions of the development of large scale structure and galaxy formation.  It is exquisitely well established from the observation of the dipole temperature variation of the Cosmic Microwave Background that our Galaxy has a deviant motion of 630 $\kms$  \citep{1996ApJ...473..576F} but there is a very poor understanding of how that arises.  There are local components like our collision course toward M31\citep{2007JCAP...10..016H}, the repulsion by the Local Void, and the pull from the Virgo Cluster and surrounding structure \citep{2008ApJ...676..184T}.   On a larger scale, it is now apparent that the Galaxy lies at the periphery of an overdense region centered near the giant clusters in Norma, Hydra, and Centaurus at distances of 3000$-$5000 $\kms$.  It has been long established that we are part of a flow in that direction  \citep{1986ApJ...307...91L, 1988ApJ...326...19L}.  However there has been a lively debate about whether the flow ends there, or continues on to the Shapley Concentration at 15,000 $\kms$  \citep{2005ApJ...635...11P, 2006ApJ...645.1043K, 2006MNRAS.368.1515E,  2010ApJ...709..483L, 2011arXiv1102.4356B} , or continues even further \citep{2010MNRAS.407.2328F, 2010ApJ...712L..81K}.  There are important things to be learned from cosmic flows.

The discussion will begin with a description of current observational material, with brief mention of anticipated developments.  This section will be followed by a discussion of Wiener Filter reconstruction of  three-dimensional velocities and a density field compatible with the observations within the limitations of a linear analysis.  The viability of the Wiener Filter constrained realizations is checked in the next section with mock catalog comparisons.  A constrained simulation that provides a model that is in general but not fine-tuned accordance with the local structure provides the test.  It will be demonstrated that the procedure does a good job of recovering the observed structure and velocity field in the domain of observations and does a reasonable job of anticipating the structure at distances well beyond the observations.

\section{Data}
\label{sec:data}

This article mixes observations and theoretical modeling.  Observed distances are based on a generally accepted zero-point scale \citep{2001ApJ...553...47F} so are reported in Mpc.  The natural distance unit for the models is $h^{-1}$ Mpc where $h$ expresses the Hubble Constant H$_0 = 100h$ km/s/Mpc.  Distances related to interpretation of models are often quoted as $h^{-1}$ Mpc  and $h = 0.74$ gives an appropriate translation to the observations.  If a distance is not tagged with the parameter `$h$' then it is on the observed scale.

\subsection{Observed peculiar velocities} 

In  \citet{2008ApJ...676..184T} we published a catalog of distances that will be hereafter identified as "Cosmicflows-1"\footnote{Available in the Extragalactic Distance Database \citep{2009AJ....138..323T} http://edd.ifa.hawaii.edu}.  The catalog contains distances for 1797 galaxies within 3300~km/s, drawing from four different methods: the Cepheid Period-Luminosity relation, Tip of the Red Giant Branch and Surface Brightness Fluctuation luminosity indicators, and the galaxy luminosity-rotation rate or Tully-Fisher correlation. The first three methods give distances with accuracies of 10\%, but in the case of the first two only to very local galaxies, and in the case of the third only to early types that are preferentially located in high density regions.  The Tully-Fisher method, invoking the correlation between the neutral HI gas rotation rate and the integrated magnitude of a spiral galaxy, has an $rms$ accuracy of 20\% with measurement in a single band but the lesser accuracy per measurement is offset by the availability of large samples widely distributed over the sky and in distance.\\

Peculiar velocities ($V_{pec}$ in \kms) and corresponding errors are computed from distances ($d$ in Mpc) and their errors assuming a value for the Hubble constant H$_{0} = 74$ km/s/Mpc \citep{2008ApJ...676..184T}:
\begin{equation}
V_{pec}=V_{CMB}-d* {\rm H}_{0}  \hfill ~(\kms)\\
\end{equation}
\begin{equation}
\Delta V_{pec}=-\Delta d* {\rm H}_{0}  \hfill ~(\kms)\\
\end{equation}

\begin{figure*}
\begin{tabular}{ll}
\includegraphics[width=0.4\textwidth]{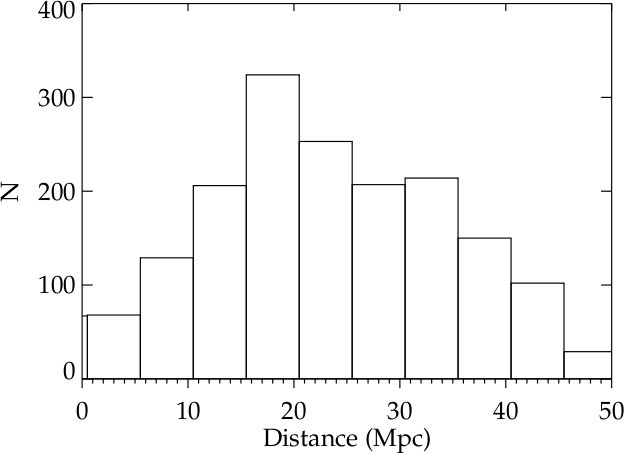}&
\includegraphics[width=0.40\textwidth]{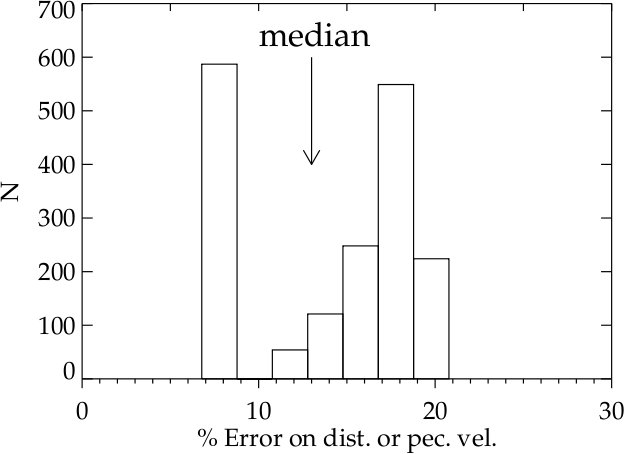}\\
\includegraphics[width=0.40\textwidth]{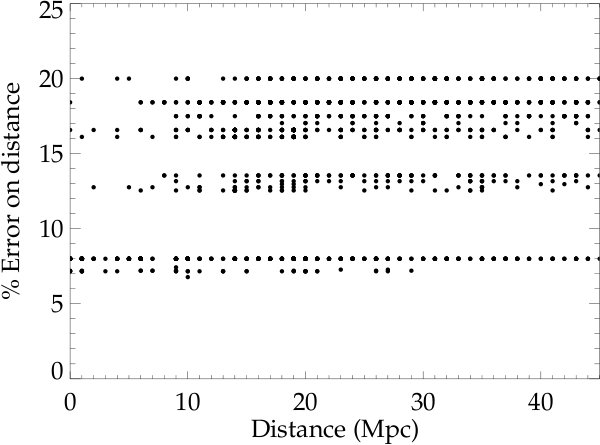}&
\includegraphics[width=0.4\textwidth]{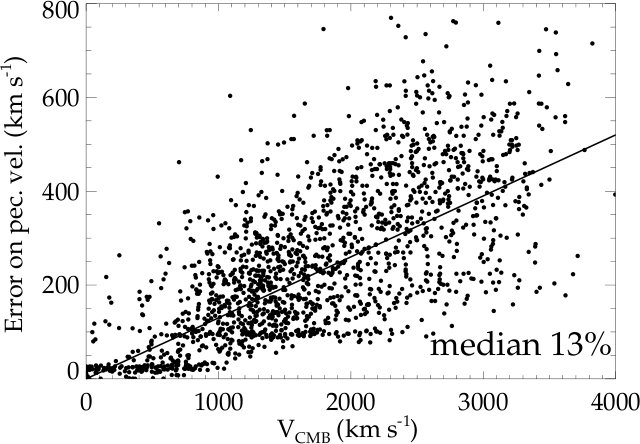}
\end{tabular}
\caption{
Observational data in Cosmicflows-1 catalog. {\it Top left.}  Distance distribution of the 1,797 individual measurements within 742 groups with accurate distance measurements. {\it Top right.} The distribution of \% errors in distances and peculiar velocities in the observed catalog.   The median error of the CF1 catalog is 13\% in both distance and peculiar velocity. {\it Bottom left.} Percent error in distance as a function of distance.  {\it Bottom right.}  The absolute errors on peculiar velocities increase with the depth of the catalog. 
}
\label{data-CF1}
\end{figure*}

The distribution of {\it rms} percent errors in distance is shown in Figure~\ref{data-CF1}.  The peak at 20\% results from single measures with the Tully-Fisher method.  Distance measures with uncertainties larger than 20\% are not accepted as sufficiently accurate and thus are not used in this catalog.  As seen on Figure~\ref{data-CF1}, the median {\it rms} error in both distance and peculiar velocity is 13\%.

\begin{figure*}
\begin{centering}
\begin{tabular}{c}
\includegraphics[width=0.9\textwidth]{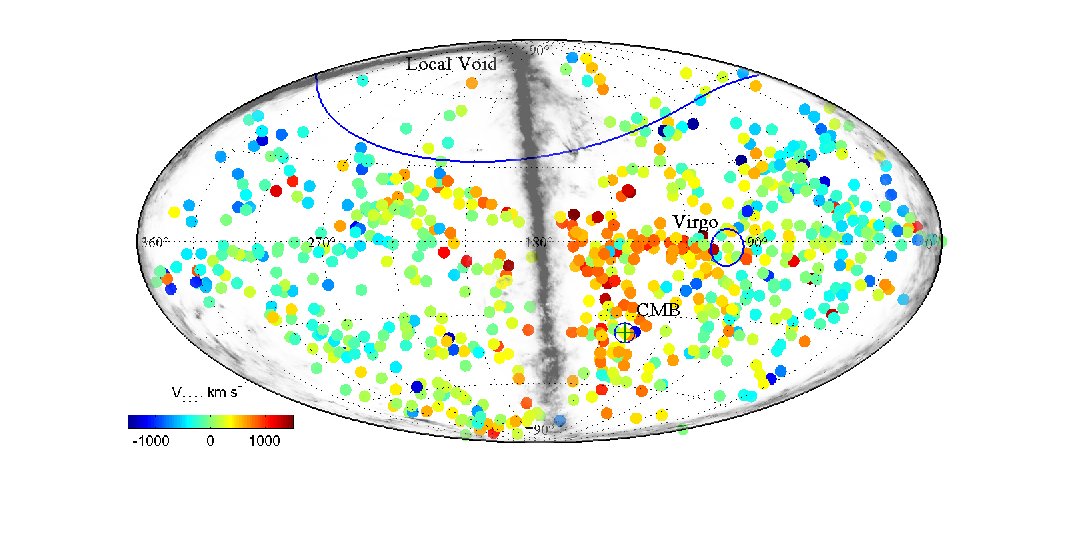}\\
\includegraphics[width=0.6\textwidth]{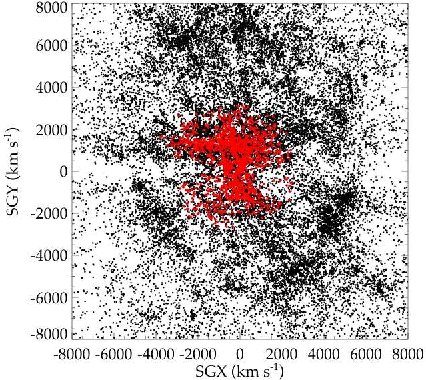}
\end{tabular}
\end{centering}
\caption{Top: the V8K redshift catalog containing 30,126 redshifts displays the structure in the distribution of galaxies within a box of radius 8000~km/s (black dots).  The distribution of 1,797 galaxies with measured distances in Cosmicflows-1 is superposed in red.  The Galactic plane lies at supergalactic coordinate SGY=0 and accounts for a band deficient in sources. Bottom: an Aitoff all sky projection of the 742 grouped peculiar velocities in Cosmicflows-1 with galactic extinction over-plotted in grey shading.
}
\label{maps}

\end{figure*}

It is to be appreciated that the Cosmicflows-1 catalog is not a complete sample. Incompleteness does not affect the reconstruction of the three-dimensional velocity field as long as the underlying individual distances are not subject to biases in the distance measures.\footnote{The subject of biases is discussed in detail by \citep{2000ApJ...533..744T}.} The observed peculiar velocity field can be sparse in an inhomogeneous way without leading to deviant results in the reconstruction as it will be demonstrated later in this paper.  Figure~\ref{maps} 
~shows the distribution of galaxies with distance measures in the Cosmicflows-1 catalog in supergalactic coordinates SGL, SGB. The velocity distribution of the sample is shown in Figure~\ref{data-CF1}.

The general features of the large scale structure within 8000~km/s are illustrated in Figure~\ref{maps}.  
~This figure is a representation of the 30,124 galaxies in the redshift catalog drawn from the literature that we call `V8K'\footnote{The V8K catalog is available at http://edd.ifa.hawaii.edu}.   
The  1,797 data points with measured peculiar velocities within 3,000 $\kms$ are included in this larger catalog. The WF is sensitive to structure on scales larger than the domain of the distance measurements.  The V8K catalog is used in this paper to compare the observed and derived densities from small to large scales\footnote{Alternatively, comparisons could be made with 2MRS, the 2MASS Redshift Survey  \citep{2005IAUS..216..170H} which extends farther.  The two catalogs are qualitatively similar but V8K includes fainter galaxies so provides denser coverage.}.

The catalog of distances is a work in progress.   Data relevant to the measurement of distances are being accumulated in the Extragalactic Distance Database, and an extended version of this catalog "Cosmicflows-2" is in preparation.  Observations are being extended to $\sim 15,000$~km/s in a program of HI observations and photometry that we are calling "Cosmic Flows" \citep{2011MNRAS.tmp..466C}.  HI observations are being made in the context of the Cosmic Flows Large Program with the 100m Green Bank Telescope at the National Radio Astronomy Observatory and complementary southern observations with the Parkes Telescope in Australia.  HI profiles for more than 13,000 galaxies have been measured consistently by the same pipeline \citep{2009AJ....138.1938C}.  The photometry for the program is being acquired from the University of Hawaii 2.2m Telescope \citep{2011arXiv1103.5106C}, the multiband survey Pan-STARRS in the North and, it is anticipated, SKYMAPPER in the South.



\section{Analysis: Theoretical Tools}
\label{sec:analysis}

The analysis of the velocity field is performed here within the standard model of cosmology. The model assumes that structure has evolved out of a primordial Gaussian random perturbation field via gravitational instability. This hypothesis implies that on large enough scale the flow is curl-free and that the velocity and gravitational fields are linearly related. The Bayesian Wiener Filter Constrained Realization (WF/CR) analysis is performed here within the standard $\Lambda$CDM model assuming the cosmological parameters derived from the WMAP5 data base \citep{2009ApJS..180..330K}.  The method is tested against mock catalogs drawn from a constrained simulation run within the $\Lambda$CDM/WMAP3 framework.

\subsection{Wiener Filter and Constrained Realizations}
\label{sec:wf-cr}

Peculiar velocity surveys constitute a very noisy and sparse data set diluted across the whole sky.  Nonetheless, the peculiar velocities directly sample the gravitational field induced by the large scale structure (LSS) of the universe and therefore can be used to uncover the underlying density and velocity field. The WF algorithm provides the optimal tool for the reconstruction of the LSS from peculiar velocity surveys. The general WF  framework has been thoroughly reviewed in \citet{1995ApJ...449..446Z} and \citet{2009LNP...665..565H}. The application of the WF  to peculiar velocity observations was first presented in \citet{1999ApJ...520..413Z}. 

In the standard model of cosmology the LSS emerges out of a primordial random Gaussian perturbation field via gravitational instabilities. In the linear regime of deviations from the homogeneous and isotropic Friedmann expansion the density and velocity fields retain their Gaussian nature. The WF constitutes the optimal linear  minimal variance estimator given a general data set and an assumed prior cosmological model. It can be shown that in the case of an underlying Gaussian random field the WF coincides with the optimal Bayesian estimator of the field given the data \citep{1995ApJ...449..446Z}. 


The linearity assumption implies that the density and velocity fields are reconstructed as if the linear regime is valid today on all scales. However, the present epoch LSS is non-linear, with a smooth transition from the very large scales where the linear theory prevails down to small scales that are governed by non-linear dynamics.  The scale of transition from the linear to the non-linear regime depends on the nature of the field under consideration. The scale of non-linearity of the velocity field is roughly a few Mpc  and for the density field it is about a factor of ten larger. It follows that the WF  algorithm provides an excellent tool for the reconstruction of the present epoch velocity field.

A final comment on a property of
 the WF reconstruction is due here. The WF constitutes a conservative estimator which is designed to find the optimal balance between the observational data and the prior model one is assuming in constructing the WF \citep{1995ApJ...449..446Z}. In regimes where the data is dense and accurate, the WF  is dominated by the data while in regimes where the data is very noisy or very sparse the WF recovers the prediction of the prior model. In the standard model of cosmology the expected mean density departures and velocity fields are zero. In observational surveys of peculiar velocities the statistical error of individual measurements increases linearly with distance and the data sparseness increases with the distance. The degradation in the quality of the data with the distance implies that the WF reconstruction is expected to attenuate the reconstructed fields towards the null fields.  

The WF presents the conditional mean field given the data and the assumed prior model.  The variation of the actual field around the WF mean field can be sampled by performing CRs of the underlying Gaussian density field  \citep{1987ApJ...323L.103B, 1991ApJ...380L...5H, 1999ApJ...520..413Z}. The scatter of the CRs provides a measure of the statistical uncertainties of the WF reconstruction.

\subsection{Decomposition into Tidal and Local Flow Fields}
\label{sec:decompose}

An interesting question given attention by modelers of the velocity field concerns the identification of the sources causing the motion. What cosmological objects such as clusters, superclusters and voids, are affecting the velocity field within a given region of space? 

For a given reconstruction of the 3D velocity field in a given region of radius $R$, the velocity field can be decomposed into two components, one that is induced by the mass distribution within the volume  and one induced by the outside  mass. This decomposition can be easily done in the linear regime \citep{2001astro.ph..2190H} where the velocity field is proportional to the density field and therefore the velocity and density fields are related by a Poisson-like equation:
\begin{equation}
\label{eq:ql1}
 \nabla \cdot {\bf v}= -  H_0 f(\Omega_m, \Omega_\Lambda)   \delta . 
\end{equation}
Given a density field within a given volume and assuming boundary conditions on that volume, the Poisson-like equation can be solved and the velocity field can be calculated. This amounts to the particular solution, to which one can always add the homogenous solution of the source-free equation, namely the Laplace equation. The consequence is the separation of the velocity field inside a given volume into its local component, namely the one induced by the internal mass distribution, and the tidal component \citep{2001astro.ph..2190H}.  It is to be noted that we used the term 'tidal' to represent the full velocity field that is induced by the external mass distribution, including the dipole term (i.e. the bulk motion of the volume due to external causes at any depth).

The WF reconstructs  the density and  velocity fields  on a grid that spans some computational  box.   Alternatively, one can use the WF velocity field and calculate the linear density field by taking the divergence of the velocity field. Suppose that the velocity field is to be separated into the component induced by the density within a sphere of radius $R$, enclosed within the box, and the residual of that component. This separation is done by solving Eq. \ref{eq:ql1} for a density field set to the WF field within $R$ and taken to be zero outside $R$. In practice the operation is carried out using an FFT solver and imposing zero-padding outside the desired volume. The resulting field is the divergent  or local component, and the residual of the full WF velocity field after vector subtraction of the local component constitutes the tidal field induced by the mass distribution beyond the sphere of radius $R$. Visual inspection of the local field often suffices to infer the existence of "great attractors" that lie outside the sphere. A multipole decomposition of the tidal field provides a more quantitative characterization of the tidal field.

\subsection{Structure Outside the Data Zone}
\label{sec:outside}

Peculiar velocity  data sets consist of a finite number of data points taken within a finite volume, the data zone. Peculiar velocities have a much larger correlation length than the underlying density field. It follows from the latter fact that the WF reconstruction can be applied on a volume larger than the data zone.  The extent and quality of the reconstruction depends on the assumed prior model, i.e. the power spectrum, and the quality of the data:  the magnitude of the errors and the sparseness and the space coverage of the data. In the following, the WF reconstruction is to be conducted within two boxes of size $L=64$ and $160 \hmpc$, centered on the origin of the Supergalactic coordinates system.

\subsection{Testing with Mock Catalogs}
\label{sec:mock}

A detailed analysis of the tests conducted with mock catalogs is available in the accompanying paper (Hoffman et al. 2011).  We give an overview here.
A mock velocity catalog has been drawn from a constrained simulation of the local universe. The simulation was performed within the CLUES project\footnote{http://www.CLUES-project.org}, specifically the case discussed by \citet{2011ApJ...726L...6C}.  The simulation is contained within a box of $L=160\hmpc$, hereafter BOX160, assuming the WMAP3 cosmological parameters. This model is a revision of the \citet{2003ApJ...596...19K} constrained simulation, with a different  realization of the random field and more modern estimation of the cosmological parameters.  A visualization of the simulation and some discussion of the structural properties is presented in \citet{2011ApJ...726L...6C}. 
The simulation reproduces the main players of the local cosmography. In particular it recovers a representation of the Local Supercluster, the Virgo and Coma clusters, the Great Attractor and the  Perseus-Pisces Supercluster. It also reproduces an environment like the neighborhood of the Local Group (LG) and its cold Hubble flow. The simulation provides  a very good dynamical replication of the local universe and thereby constitutes a good `laboratory' for tests of our tools of analysis.

This is a dark matter only simulation. A hierarchical friends-of-friends halo finder was applied to the simulation and a halo catalog has been constructed. A mock galaxy catalog  has been constructed using the algorithm of the conditional luminosity function \citep{2007MNRAS.376..841V}. An `observer' has been placed at the position of the LG and the peculiar velocity of 1,797 galaxies has been measured, assuming a Poisson distribution of errors centered around 13\%. The selection of the galaxies is done randomly in a sphere of $R=30\hmpc$ centered on the LG in order to mimic the observed redshift distribution.

\begin{table*}
  \begin{center}
    \begin{tabular}{lcc}
        \hline\hline
        &  $V(R_g=5\hmpc) $         &    $ V(R_g=10\hmpc) $   \\
    \hline
      Slope                      &    $0.81$              &     $0.92$                            \\
      Corr. Coeff.              &    $0.89$              &     $0.94   $                          \\
      Scatter                    &    $57$ km/s        &     $35$ km/s                       \\
      Mean                    &    $9.2$ km/s       &     $3.5$ km/s                      \\
      St. Dev.                  &    $63$ km/s        &     $36$ km/s                       \\
      Skewness         &    $0.41$              &     $0.26$                             \\
      Kurtosis           &    $0.85$              &     $1.12$                             \\
    \hline
    \end{tabular}
    \vskip 0.5cm
    \caption{Comparison of  velocity fields with the WF reconstructed fields from the mock catalog.
    The velocity field is represented by the magnitude of the velocity vector. 
    The comparison is done for $R_g=5$ and $10\hmpc$.
    The first three rows present the slope, correlation coefficient and scatter of the linear regression of the simulation and the WF reconstruction.  The following rows give
    the first 4 moments (mean, standard deviation, skewness and kurtosis) of the residual of the WF reconstructed fields from
    the simulated fields.  }
    \label{table:compare}
  \end{center}
\end{table*}

\begin{figure*}
 \begin{tabular}{cc} 
\includegraphics[width=0.51\textwidth]{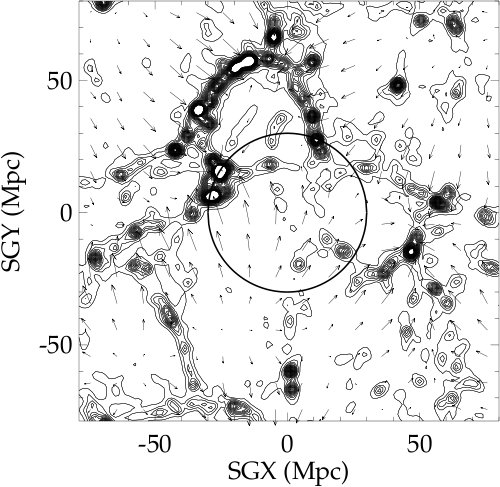}&
\includegraphics[width=0.51\textwidth]{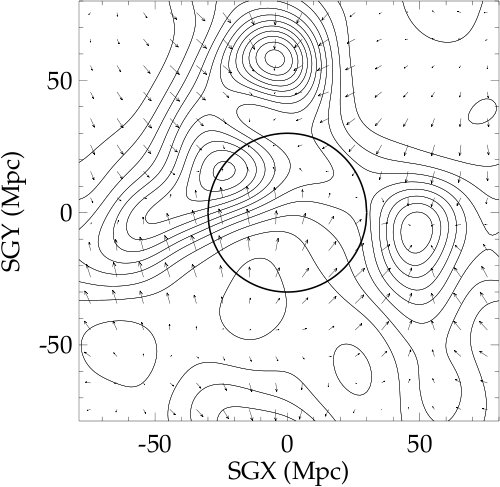}\\
\includegraphics[width=0.51\textwidth]{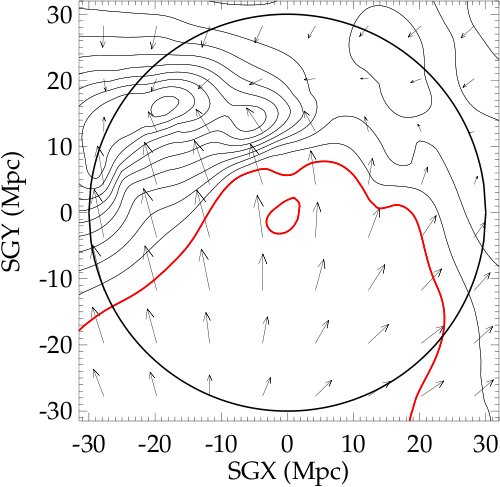}&
\includegraphics[width=0.51\textwidth]{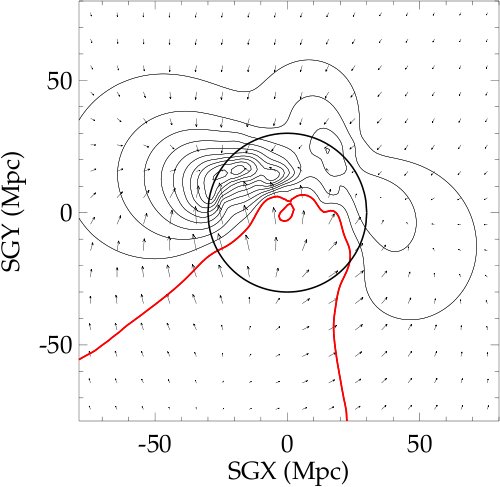}
\end{tabular}
 \caption{
Reconstruction of the mock catalog, sampled from a BOX160 constrained simulation of the local universe. The upper row shows the density and velocity fields of the simulation, smoothed at 1 (left) and 10 $\hmpc$ (right).
The circle show the zone where a mock catalog of peculiar velocities, similar to cosmicflows-1, was created (the data-zone). The lower row shows the WF reconstruction within the data-zone (left) and at a larger scale than the data-zone (right).  The contours represent the logarithm of the recovered density field normalized by the 
   mean matter density of the universe (positive densities only), and the arrows represent the recovered velocity field. 
   The length of the arrows are proportional to the amplitude of velocity. 
   Note that the reconstruction is made only from the radial peculiar velocities of the mock data drawn from the central sphere of radius $R=30\hmpc$  (the data-zone), with a poisson distribution around 15\% error.
   There is excellent agreement within the $30\hmpc$ ring between the smoothed input (top right) and WF reconstruction at the same smoothing scale (bottom panels at two scales).
}
\label{fields}
 \end{figure*}

The upper right panel of Figure~\ref{fields} shows the Supergalactic Plane of the constrained BOX160 simulation. The density and velocity fields are Gaussian smoothed with $R_g=1\hmpc$. At such high resolution the density field lies well within the non-linear regime. The WF reconstruction is applied to the simulation with alternative smoothing lengths of $R_g=5$ and $10\hmpc$. A visual inspection of the maps reveals that the reconstruction recovers the overall LSS of the box. The comparison in Table \ref{table:compare} shows that the reconstruction is best with the larger Gaussian smoothing and is excellent within the $30\hmpc$ annulus of the data zone.  The extrapolation at larger distances recovers the gross  features of the LSS but at a degraded quality. It follows that the WF can be used to provide a rough expectation of the LSS in unobserved regions.

A linear regression of the N-body and WF reconstructed fields is given in Table \ref{table:compare}. The table also provides the mean, standard deviation, skewness and kurtosis of the residual of the WF from the N-body fields.

\section{Cosmography}
\label{sec:results}

\subsection{The volume within 40 Mpc}

\begin{figure*}
\begin{centering}
\begin{tabular}{ll}  
\includegraphics[width=0.42\textwidth]{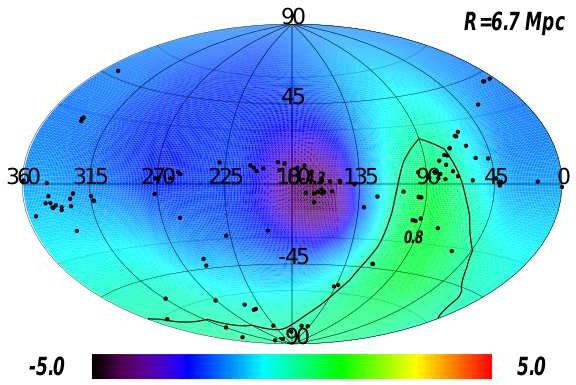}&
\includegraphics[width=0.42\textwidth]{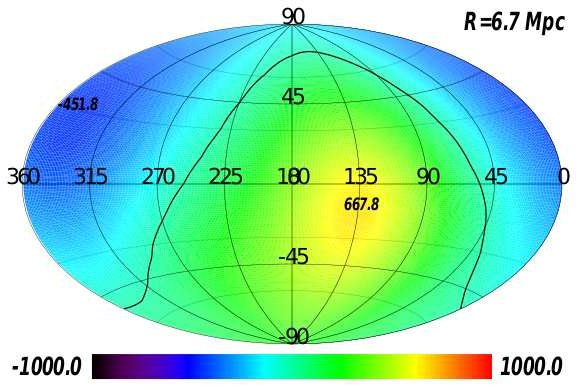}\\
\includegraphics[width=0.42\textwidth]{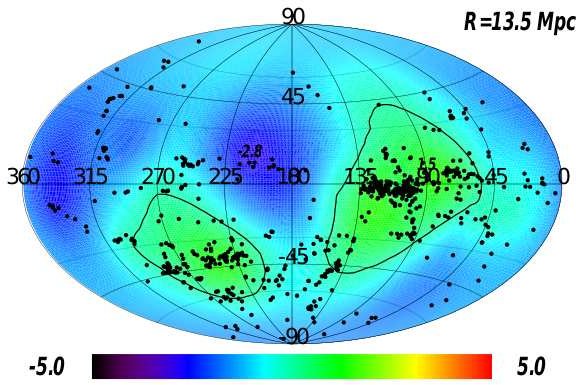}&
\includegraphics[width=0.42\textwidth]{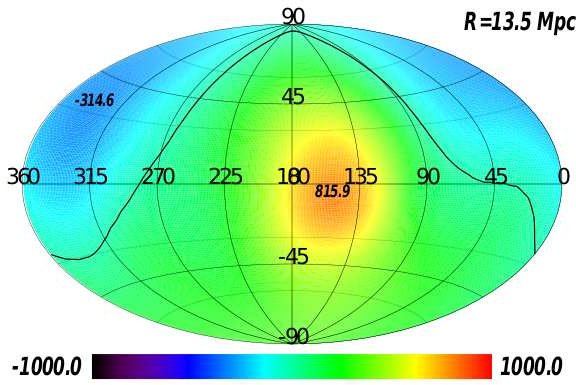}\\
\includegraphics[width=0.42\textwidth]{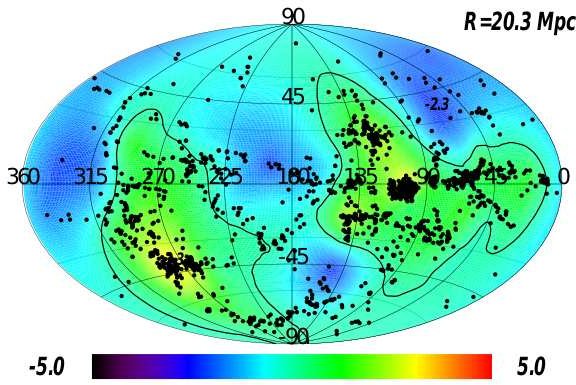}&
\includegraphics[width=0.42\textwidth]{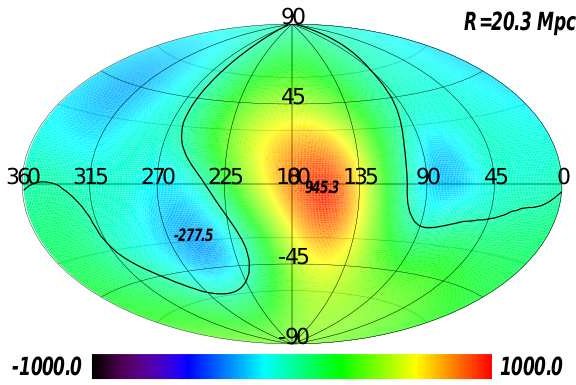}\\
\includegraphics[width=0.42\textwidth]{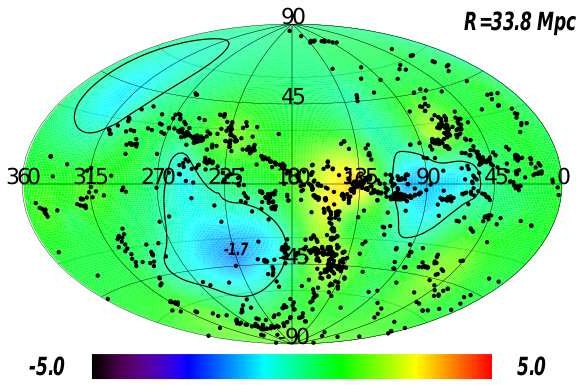}&
\includegraphics[width=0.42\textwidth]{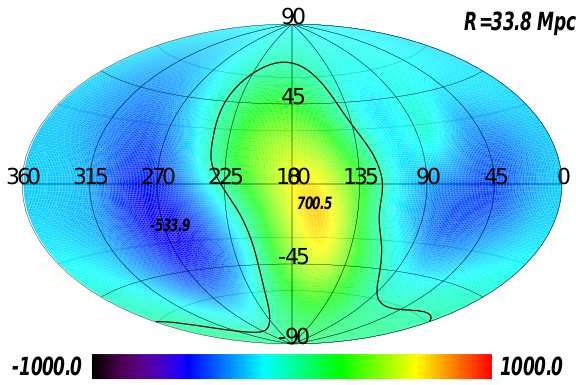}\\
\end{tabular}
\caption{ Aitoff all-sky maps in supergalactic coordinates of reconstructed densities and velocities in shells within 40 Mpc.  From top to bottom: shells centered at $5/h = 6.7$ Mpc, $10/h = 13.5$ Mpc, $15/h = 20.3$ Mpc, and $25/h = 33.8$ Mpc where $h = 0.74$.  Left column:  the color map and contours give the WF linear density field reconstructed from the Cosmicflows-1 catalog of distances.
The black dots identify galaxies from the V8K catalog within a shell of $R \pm 3.5$ Mpc. Right column: the color map gives the reconstructed radial velocity field in the same distance shells.}
\label{inside}
\end{centering}
\end{figure*}

The series of panels within Figure~\ref{inside} illustrate the broad features of the WF reconstruction within the radius containing distance measurement constraints; the data-zone.  The contours of colors illustrate the WF reconstructions based on the observed velocity field.  The left panels display linear density and the right panels display radial peculiar velocities.  In the left panels the individual points that are plotted are galaxies drawn from the V8K catalog that lie within the specific distance shells.  The pronounced correlation of galaxies with the high density WF contours provides a visual demonstration that the WF methodology can recover the 3D density distribution of the local universe from a sole knowledge of peculiar radial velocities.  It is to be noted here that the V8K galaxies are a rough proxy for the actual spatial distribution of nearby galaxies.  Galaxies are plotted based on their redshifts, assumed to be rough estimators of their distances.  The WF reconstructed radial velocity field is shown in the right panels.    

\begin{figure*}
\begin{centering}
\begin{tabular}{ll} 
\includegraphics[width=0.4\textwidth]{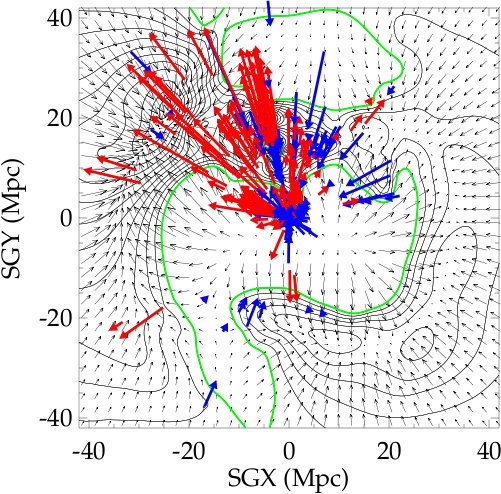} &
\includegraphics[width=0.4\textwidth]{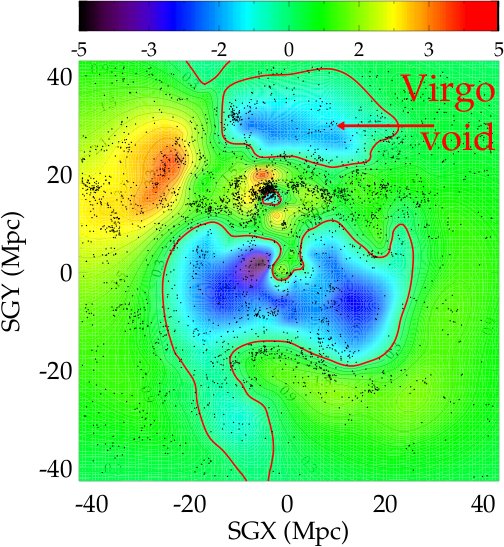}\\
\includegraphics[width=0.4\textwidth]{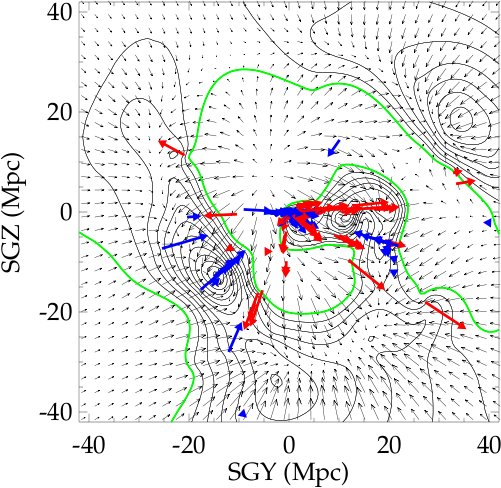}&
\includegraphics[width=0.4\textwidth]{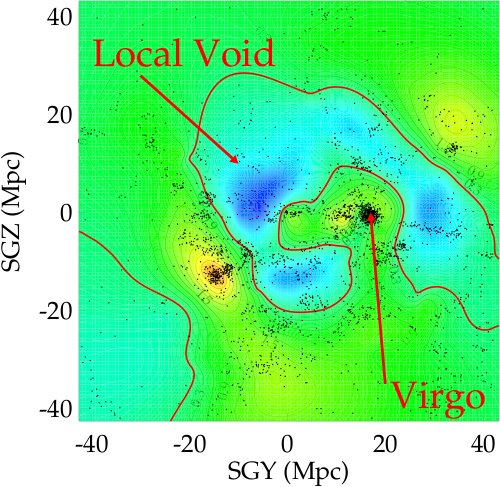}\\
\includegraphics[width=0.4\textwidth]{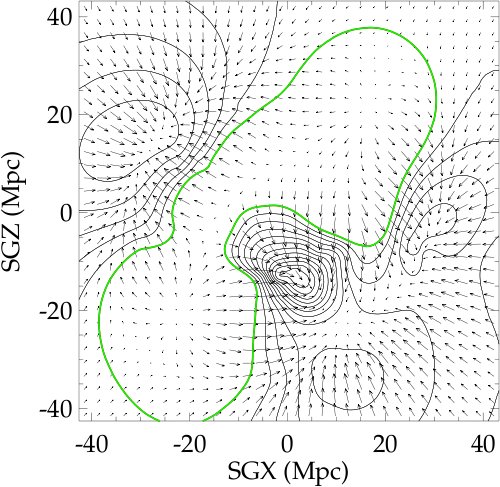}&
\includegraphics[width=0.4\textwidth]{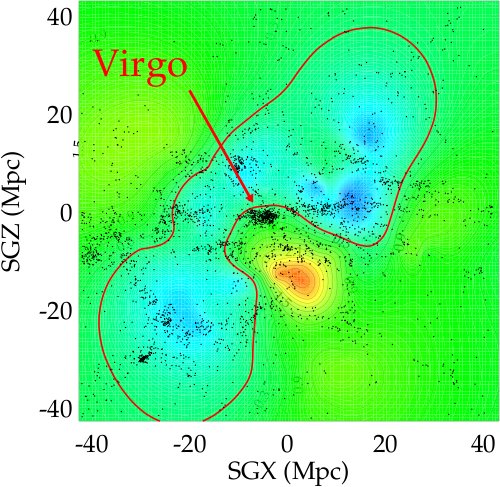}
\end{tabular}
\caption{ 
Orthogonal views of WF reconstructions of local structure.  Slices of $\pm 10$~Mpc.  SGX-SGY slice is centered at SGZ=0 and SGY-SGZ slice is centered at SGX=0 while SGX-SGZ slice is displaced to SGY=16.9 Mpc to coincide with the Virgo Cluster.  At left, over density contours are spaced at 0.3 intervals on a linear scale.  At right, colors represent over and under densities.  Individual galaxies from the V8K catalog are plotted on the right.  The local part of reconstructed 3D peculiar velocities are shown by black arrows on the left.  Observed peculiar velocities are represented by red (outward) and blue (inward) arrows in the top two panels that include our location at the origin.
}
\label{WF-CF1-slices}
\end{centering}
\end{figure*}

Figure~\ref{WF-CF1-slices} provides orthogonal views of slices of the WF reconstruction.  The panels on the left illustrate densities as contours and velocities as vectors.  Model velocities are black and on a grid and observed radial velocities are in color (red: outward; blue: inward).  The panels on the right show densities again but now with color and galaxies within the slices from the V8K redshift catalog are superimposed.  With the top two rows the slices in the normal direction are on the cardinal axes (SGZ=0 and SGX=0 respectively) while with the bottom plots the slice is displaced to SGY=16.9 Mpc to coincide with the main structure of the Local Supercluster which runs through the Virgo Cluster on a line roughly parallel to the SGX axis.  This structure is seen top-down in the upper panels and edge-on in the lower panels.


The WF reconstruction only provides a description of the linear regime and hence it has a limited power to reconstruct the present epoch structure on small scales.  However, the WF is shown to be very good for a recovery of the important large scale features.   Perhaps the most interesting revelation with the current study is the nature of the under dense region that wraps around the Local Supercluster centered on the Virgo Cluster (SGY=17 Mpc near SGX = SGZ = 0).  The larger part at SGZ positive and $-20 < {\rm SGY} < +10$ Mpc is the Local Void \citep{1987nga..book.....T, 2008ApJ...676..184T}.   There is also a pronounce void directly behind the Virgo Cluster at SGY $\sim +30$ Mpc and SGX $\sim 0$ that we will call the `Virgo Void'.  It becomes evident here (best seen in the middle panels of Fig.~\ref{WF-CF1-slices}) that the Local and Virgo voids are linked by a generally under dense region that raps above the Local Supercluster at SGZ $\sim 15$ Mpc.  Sparse filaments do course through this volume \citep{1987nga..book.....T}.  However the overriding aspects of our local region are, first of all, the general {\it under} density across a region of $\sim 60$ Mpc but, secondly, the bridge of {\it over} density of the Local Supercluster at SGY $\sim 18$ Mpc, SGZ $\sim 0$ and extending to negative SGZ, and a width in SGX that establishes links to major structures including the Centaurus and Hydra clusters (seen best in the top panels of Fig.~\ref{WF-CF1-slices}).  

These features are seen in other reconstructions of the local neighborhood \citep{1996MNRAS.283..367D, 1999MNRAS.308....1B, 2006MNRAS.373...45E}.  \citet{2007MNRAS.382....2R} emphasize the important dynamic role of voids.  \citet{2009MNRAS.400..183K} provide hints of voids on scales of 150 Mpc.  The present study is distinguished by a higher density of distance information on a smaller scale and, in particular, benefits from the observation of the kinematic signature of void expansion; the discontinuity in peculiar velocities between galaxies in the local wall of the void and galaxies beyond the wall  \citep{2008ApJ...676..184T}.  The same distance constraints were used in the study by \citet{2010ApJ...709..483L} but only to confirm the properties of the bulk flow of the local region within 3000~\kms.

\begin{figure*}
\begin{centering}
\begin{tabular}{c}  
\includegraphics[width=0.42\textwidth]{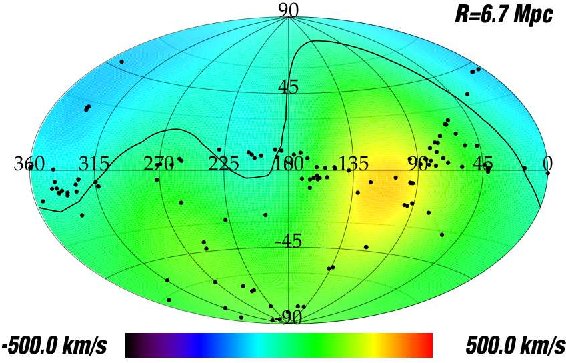}\\
\includegraphics[width=0.42\textwidth]{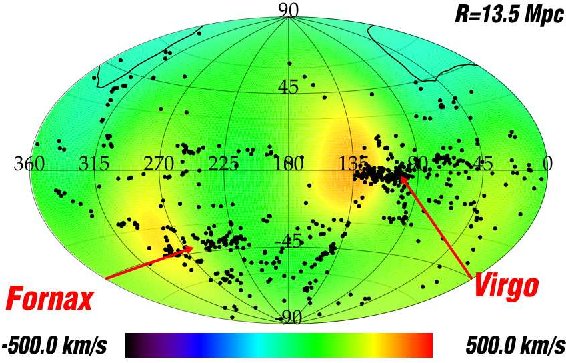}\\
\includegraphics[width=0.42\textwidth]{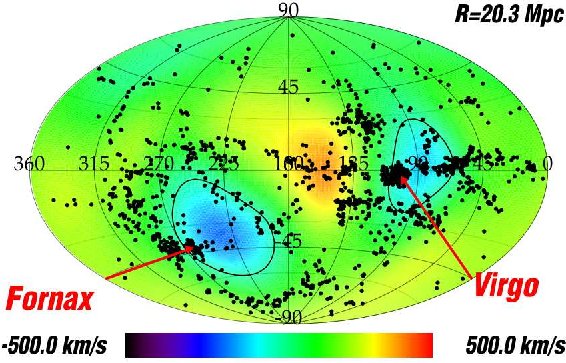}\\
\includegraphics[width=0.42\textwidth]{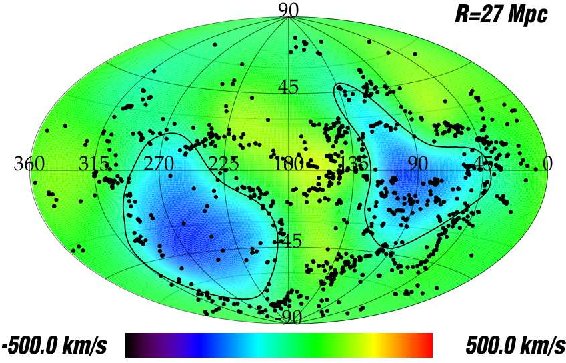}\\
\end{tabular}
\caption{ Aitoff all-sky maps in supergalactic coordinates of the radial part of the reconstructed divergent velocities in shells within 40 Mpc.  From top to bottom: shells centered at $5/h = 6.7$ Mpc through $30/h = 40$ Mpc where $h = 0.74$.  The color map and contours give the WF divergent field reconstructed from the Cosmicflows-1 catalog of distances.
The black dots identify galaxies from the V8K catalog within shells of $R \pm 3.5$ Mpc. }
\label{aitoff-div}
\end{centering}
\end{figure*}

The series of panels in Figure \ref{aitoff-div} present the evolution of the local field on Aitoff all-sky projections at increasing distances.
The plotted velocity field is the radial part only of the local field associated with mass within 40 Mpc. The top panel at 6.7 Mpc shows a motion from Supergalactic north to south and toward the direction of the Virgo Cluster, which is beyond this slice. The panels at 13.5 and 20.3 Mpc display the front fall away from us toward the Virgo Cluster and backside infall toward us. The same effect is seen around the Fornax Cluster: a front and a backside infall around sgl=270, sgb=$-45$. The color bar gives the velocities in km/s. Within this local volume the largest motion is toward a location at sgl=150, sgb=0. The farthest shells show the fading of the WF reconstruction towards the null fields as it should from construction.

\subsection{Extrapolation outside the data-zone: the Great Attractor}

The WF  reconstruction has two modes of operation. On the one hand it interpolates the structure in between the data points, providing the optimal interpolation given the data, its errors and the underlying prior model. On the other hand the WF extrapolates the structure to regions not covered by the data  \citep{1995clun.conf..135Z, 1995ApJ...449..446Z, 2001astro.ph..2190H}. The long range coherent nature of the peculiar velocities makes the WF an effective tool to probe large scale structure beyond the spatial confines of the data.
In the current situation, the catalog of observed peculiar velocities extends to 3,000 $\kms$.  In order to probe to larger scales the WF has been applied to a $160\hmpc$ box, centered on the Local Group, on a $128^3$ grid.  

\begin{figure*}
\begin{tabular}{cc} 
\includegraphics[width=0.42\textwidth]{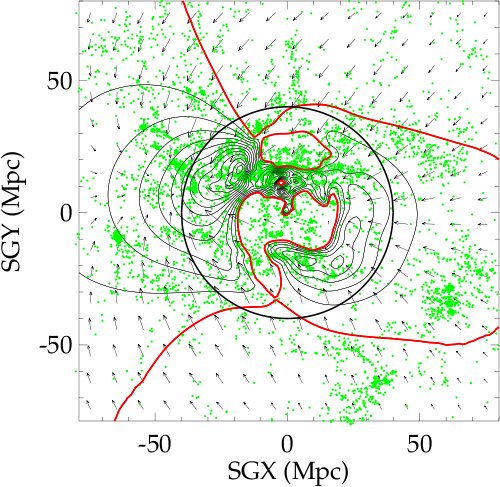}&
\includegraphics[width=0.42\textwidth]{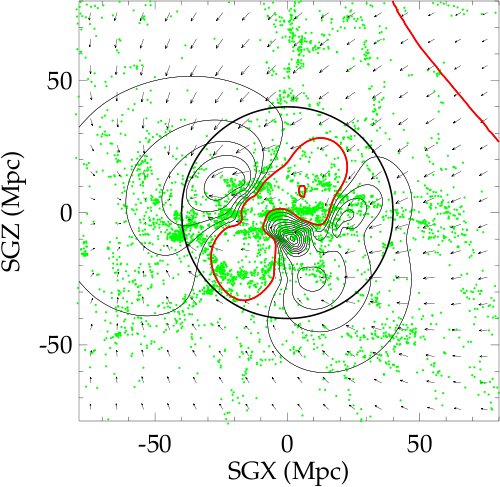}\\
\includegraphics[width=0.42\textwidth]{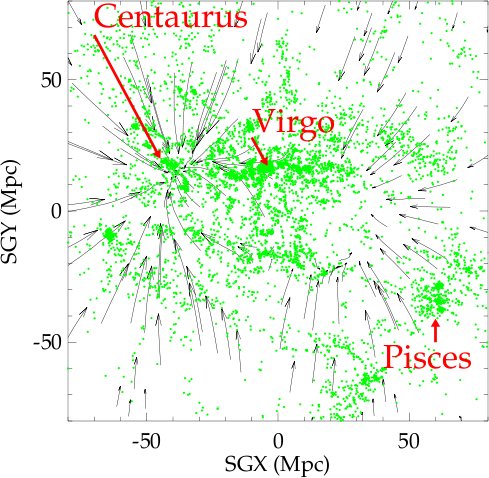}&
\includegraphics[width=0.42\textwidth]{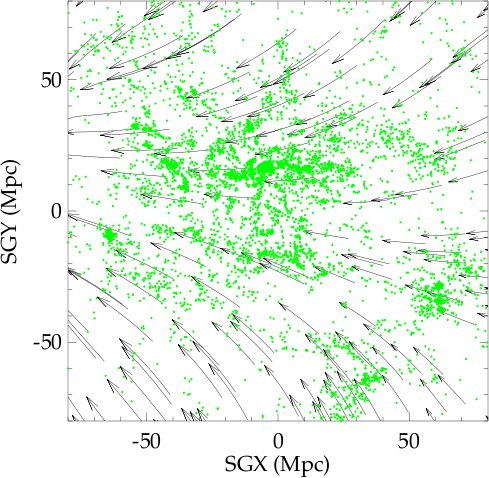}\\
\end{tabular}
\caption{ 
Comparison of the reconstruction with the actual galaxy distribution in the Local Universe at a larger scale.
Top row: WF reconstructed density contours and full velocity vectors evaluated in a box of $\pm 80$ Mpc sides. The two slices are $\pm 10$~Mpc thick with the top left panel centered at SGY=0 and the top right panel centered at SGY=16.9 Mpc, the SGY depth of the Virgo Cluster.  
The green dots are galaxies from the V8K redshift catalog with the slices. The over density contour spacing is $0.3$ plotted in linear scale.  The mean density contour is plotted as a red solid line.
Bottom row: the reconstructed WF velocity field is decomposed into its local (left) and tidal (right) components where the local velocities are a response to the WF density reconstruction within 80 Mpc.
The decomposed velocity field is represented by flow lines. Note that the local component shows the expected  close correspondence with the density field, with flow converging to the high density peaks.
}
\label{outside}
\end{figure*}

Results on this larger scale are illustrated in Figure~\ref{outside}.  The top left and right panels are expanded scale versions of the top and middle panels of Fig.~\ref{WF-CF1-slices}.  As was described in Section 3, the WF density reconstruction degrades to the mean density on scales beyond the data-zone.  The local component of the reconstructed velocity field is dominated by the repulsion of the voids and flows toward the densest structure.  The dominant convergence at SGX $\sim -40$ Mpc, SGY $\sim +20$ Mpc is coincident with prominent structure in Norma, Hydra, Centaurus, and Pavo-Indus, the principal components of what came to be called the `Great Attractor' \citep{1984ApJ...280..470S, 1987ApJ...313L..37D}.
Consideration of the tidal part  gives the visual impression that the large scale flow on a scale of 80 Mpc is dominated by attractions outside the computation box.

 \begin{table*}
\begin{center}
\begin{tabular}{lrrrrr}
 \hline
   &    $V\ [\kms]$  &  $l_{gal}$   & $b_{gal}$  & sgl  & sgb  \\
\hline
\\
$V_{\rm bulk}$       & $378$    &  $293^\circ$  &   $18^\circ$ & $156^\circ$ & $-21^\circ$ \\
$\lambda_1 \times R$ &  $261$   &   $233^\circ$  &   $2^\circ$ & $124^\circ$ & $-80^\circ$ \\
$\lambda_2 \times R$ &  $-26$   &   $144^\circ$  &   $0^\circ$ & $1^\circ$  & $-7^\circ$  \\
$\lambda_3 \times R$ &  $-235$  &   $231^\circ$  &   $-88^\circ$ & $270^\circ$ & $-8^\circ$  \\ 
\\
 \hline
 \end{tabular} 
 \caption{
 Fitting the WF tidal field by a bulk  and shear model as in Equation \ref{eq:shear}: 
 The amplitude, Galactic $(l, \ b)$ and Supergalactic (sgl,sgb) direction of the bulk velocity and the eigenvalues and eigenvectors of the shear tensor are presented.
 The eigenvalues are multiplied by $R=30\hmpc$. 
  }
 \label{table:dipole}
 \end{center}
 \end{table*}

\begin{figure*}
\begin{tabular}{cc} 
\includegraphics[width=0.42\textwidth]{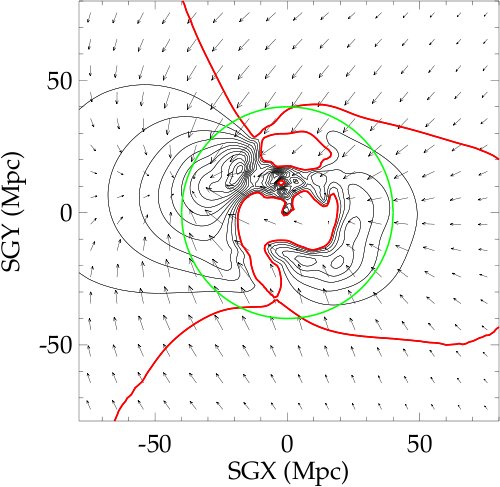}&
\includegraphics[width=0.42\textwidth]{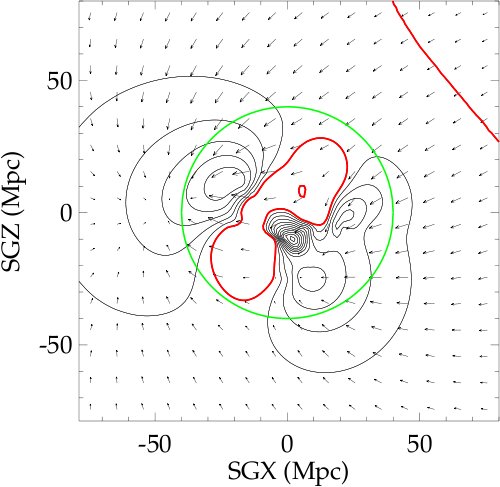}\\
\includegraphics[width=0.42\textwidth]{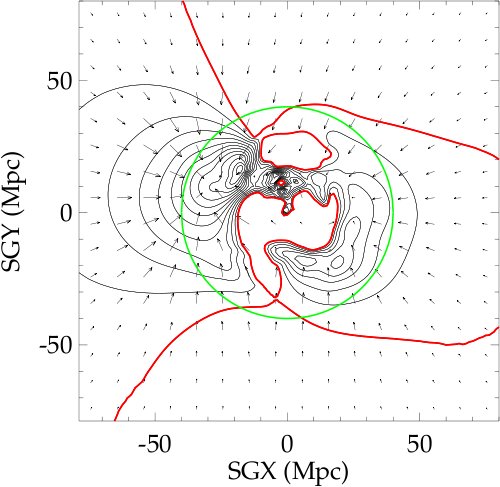}&
\includegraphics[width=0.42\textwidth]{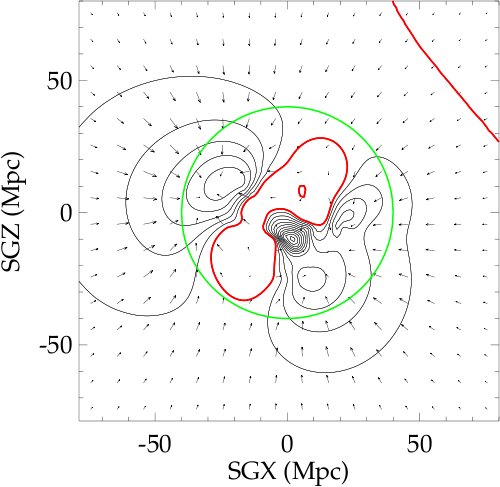}\\
\includegraphics[width=0.42\textwidth]{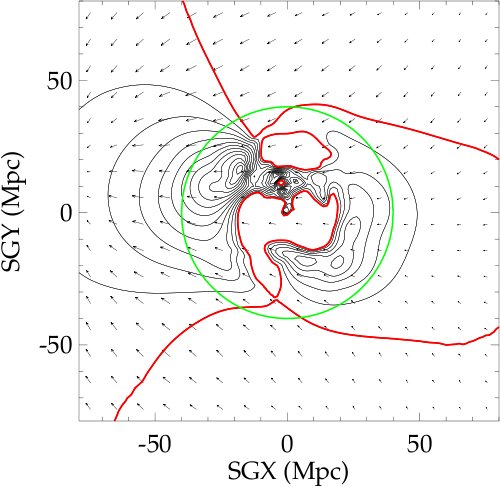}&
\includegraphics[width=0.42\textwidth]{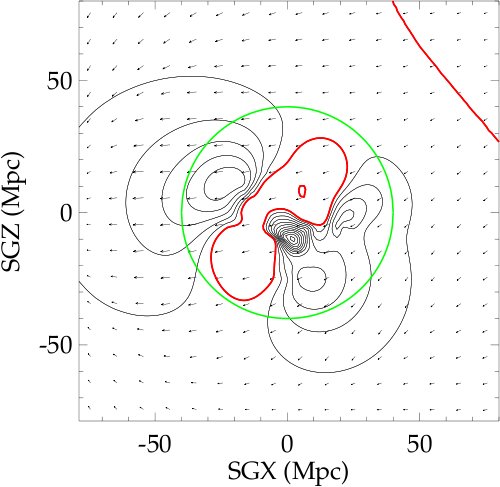}\\
\end{tabular}
\caption{ 
Separation of the WF velocity reconstruction into local and tidal components in two projections.  At left, an SGX$-$SGY plane at SGZ=0; at right, an SGX$-$SGZ plane at SGY=16.9 Mpc (coincident with the structure containing the Virgo Cluster).  On top, the full velocity reconstructions.  These vectors are decomposed into (middle) the local components induced by mass within 80 Mpc and (bottom) the tidal component, the residual after subtraction of the local velocities from the full velocities.  The concentric circles delineate the 40 Mpc extremities of the data-zone. 
}
\label{outside2}
\end{figure*}

Alternative views are given in Figure~\ref{outside2}.  The top panels are the same as with Fig.~\ref{outside} minus the superposition of the V8K galaxies.  The lower panels show the gridded local and tidal decompositions respectively, with a decomposition radius of 80 Mpc.  The choice of a decomposition radius is very large.  It comfortably includes the canonical Great Attractor region, located in the region of velocity convergences in the middle panels.  Even with this large decomposition radius the tidal flow shows a convergence of the flow just beyond this radius, thereby indicating the existence of structure beyond the Great Attractor.  It also suggests that the reconstructed Great Attractor backside infall arises from the lack of data on this scale and is not physical.


The hypothesis that the tidal velocity field is predominantly induced by a Great Attractor convergence can be tested.
The tidal velocity field has been fitted by the bulk and shear flow model of \citet{2001astro.ph..2190H}.  Consider Eq. 6 from that paper:
\begin{equation}
v_\alpha(\vec{r}) = B_\alpha + (\tilde{H}\, \delta_{\alpha\beta} +
\Sigma_{\alpha\beta}) r_\beta 
\label{eq:shear}
\end{equation}
where $B_{\alpha}$ is the bulk velocity, $\Sigma_{\alpha\beta}$ is the traceless shear tensor, $\tilde{H} \delta_{\alpha\beta}$ is a possible local isotropic perturbation, and the indices $\alpha$ and $\beta$ range over the 3 Cartesian coordinates.  The bulk and shear correspond to the dipole and quadruple terms of potential theory. The isotropic term $\tilde{H} \delta_{\alpha\beta}$ vanishes in an analysis of tides.

 \begin{table*}
\begin{center}
\begin{tabular}{llll}
 \hline
   &    $V_{\rm bulk} [\kms]$  &  $l_{gal}$   & $b_{gal}$  \\
\hline
\\
WF       & $415$    &  $295^\circ$  &   $19^\circ$  \\
CRs     &  $401 \pm 19 $   &   $299^\circ \pm 6$  &   $24^\circ \pm 4$ \\
RANs  &  $341 \pm 130$  &           &                        \\
\\
 \hline
 \end{tabular} 
 \caption{
Statistics of the bulk velocity of the full velocity field.  Row 1: magnitude and direction of $V_{bulk}$ in the WF reconstructed velocity field. Row 2: The mean and standard deviation of $V_{bulk}$ taken over an ensemble of 16 CRs.  Row 3: The mean and standard deviation of $V_{bulk}$ taken over 16 random, unconstrained realizations.  Note that for the random realizations the direction of the bulk velocity is random, hence its mean and scatter are not shown.
  }
 \label{table:CR}
 \end{center}
 \end{table*}

Table \ref{table:dipole} presents the bulk velocity, $V_{\rm bulk}$, the three eigenvalues of the shear tensor, $\lambda_1,\ \lambda_2$ and $\lambda_3$, and the direction of their corresponding eigenvectors. The eigenvalues are multiplied by $R=30\hmpc$, so they acquire the dimensionality of velocity.
A tidal flow induced by a single attractor is expected to have:\\
a)  the eigenvector corresponding to $\lambda_1$, hereafter the first eigenvector, and the bulk velocity to be aligned;\\
b) the shear eigenvalues have the following structure:\\
 $\lambda_2 =  \lambda_3 = -\lambda_1/2$. \\
 The results of Table \ref{table:dipole} imply that the local tidal flow cannot be modeled by a single attractor. The $61^\circ$ misalignment between the first eigenvector and the bulk velocity is one indicator. Yet, the strongest evidence comes from the structure of the eigenvalues, which are closer to 
a ($\lambda, 0, -\lambda$) structure which is expected for a random field point \citep{2001astro.ph..2190H}.




\subsection{Bulk velocity}
\label{sec:bulk}

\begin{table*}
 \begin{center}
   \begin{tabular}{lrrrrrrrrl}
          \hline
                             &     $V$        &       $l_{gal}$       &       $ b_{gal}$    &   $sgl$    &    $sgb$    &    $V_x$       &        $V_y$       &      $V_z$     &     alignment $V_{CMB}$  \\     
                             &     [km/s]  &       [deg]       &       [deg]  &     [deg]       &       [deg]  &  [km/s]    &    [km/s]         &      [km/s]         &    [deg] \\
                                        \hline
\\
CMB                     &     627    &         268      &       27      &   138   &  -38     &  -364      &             330   &              -390       &       0  \\
$V_{LG}$             &     593     &        273      &       29       &   139   &  -34     &         -374     &            326     &            -324        &       5\\
TIDAL                    &   382      &        298      &      16         &   161   &   -17    &         -348      &         118      &             -106         &       30\\
LOCAL                  &   304    &         234      &      36         &    98   &   -47    &        -28         &       208        &         -220            &       30\\
VIRGO (7 Mpc/$h$)       &   29      &        277       &       75         &   101   &  -3     &           -5         &          28       &         -1                 &       49\\
                               &             &                      &                     &       &       &                       &                      &                             &        2 (w.r.t. Virgo)\\
VIRGO (10.6 Mpc/$h$) &  60       &       202       &           18      &   45   & -55      &          24         &        24          &               -49        &        60\\
                               &             &                      &                     &       &       &                      &                        &                           &          65 (w.r.t.
                                Virgo)\\
      \hline
   \end{tabular}
   \vskip 0.5cm
   \caption{
 Dissecting the Local Group motion. CMB: The CMB dipole used here to transform the data to the CMB frame.
$V_{LG}$: The LG motion  reconstructed by the WF.
TIDAL: The $V_{LG}$ component induced by the tidal field, namely the mass distribution outside a sphere of
R = 30 $\hmpc$.
LOCAL: The $V_{LG}$ component induced by the local mass distribution, i.e. within $R = 30 \hmpc$ (this is the divergent component).
VIRGO $R=7 \hmpc$: The $V_{LG}$  component induced by the mass distribution in and around the Virgo cluster, specifically within a sphere of radius $R = 7 \hmpc$ centered
on [SGX, SGY, SGZ] = [3, 10, 0] $\hmpc$.
VIRGO $R=10.6 \hmpc$: Same as Virgo $R=7 \hmpc$ but for a sphere of radius $R=10.6 \hmpc$, i.e. the LG-Virgo separation.
Columns 2-4: the absolute velocity and its Galactic (l, b) direction; columns 5-9: the
Supergalactic direction and decomposition of the velocity; column 10: the alignment with the CMB dipole. }
   \label{table-vbulk}
 \end{center}
\end{table*}

The bulk velocity of any velocity field is defined here as the volume weighted average velocity of the full 3D velocity field in a given volume. For this discussion, a sphere of radius $R=30 \hmpc$ is assumed and the velocity field is assumed to be evaluated on a regular Cartesian grid.
Table \ref{table:CR} presents the bulk velocity, $V_{bulk}$,  of the the WF reconstructed velocity field. To study the robustness of WF estimator of $V_{bulk}$ a suite of 16 CRs have been constructed, using the same observational data and prior model as in the WF reconstruction.  The construction of a CR involves the complementary use of an unconstrained realization of the underlying field. These random realizations (dubbed RANs) are used as a tool for calculating the mean and scatter of the amplitude of the bulk velocity of a random sphere in the assumed $\Lambda$CDM (WMAP5) cosmology. Both the CRs, and therefore also the RANs, are evaluated on a box of $L=640 \hmpc$ on a $128^3$ grid. This large computational box is used to avoid suppression of the residual power contributed by very large scales. 
The definition of $V_{bulk} $ used here differs from the usual definition adopted in the literature. Whereas $V_{bulk}$ is often taken to be the mean velocity of a given sample of galaxies, the bulk velocity is defined here by the mean velocity of all galaxies within a given volume.



Table \ref{table-vbulk} provides information on the sources of the motion of the LG with respect to the CMB frame of reference.  The full reconstructed velocity field yields a velocity $V_{LG}$ that is very close to the CMB dipole motion, $V_{CMB}$, with only $5^{\circ}$ misalignment.  The tidal and local components have similar amplitudes and deflections but are mutually misaligned by $60^{\circ}$.  The last two rows of Table~\ref{table-vbulk} present the values of the velocity field at the position of the LG induced by the mass distribution in spheres of radius 7 and 10.6 Mpc/$h$ centered on the Virgo Cluster.  The mass in the smaller sphere induces a motion closely aligned with the cluster but the amplitude of 29 km/s is small.  The mass in the larger sphere induces a velocity with an increased amplitude of 60 km/s but the direction is $65^{\circ}$ away from the cluster.  It follows that the Virgo Cluster plays a minor role in shaping the motion of the Local Group with respect to the CMB.   The growth of the amplitude and the shift in direction with the increase in radius of the shell around the cluster are consistent with the proposition of a dominant role of the Local Void, as discussed in the next section.

\section{Discussion and Summary}

The WF methodology has been applied to the Cosmicflows-1 data base which consists of 1,797 galaxies within 3000 km/s  with  distance measurements to reconstruct the large scale structure of the local universe. The full linear density and 3D velocity fields have been reconstructed, assuming as a prior model the standard $\Lambda$CDM model with the cosmological parameters derived from the WMAP5 data base. The simultaneous reconstruction of both the density and velocity fields has enabled a detailed and critical examination of the sources that induce the observed velocity field on scales of a few tens of Megaparsecs.  The main progress  since the first application of the WF/CR application to velocity data by Zaroubi, Hoffman and Dekel (1999) has been along three main lines: better observational data, improved testing by mock galaxy catalogs, and the employment of a tidal field decomposition.  A brief summary of our main results follows.

Maybe our  most striking finding is the dominant role of the Local Void in shaping the local velocity field.  Voids are a prominent feature of our neighborhood and the Local Void is of special note.   \citet{2008ApJ...676..184T} concluded that this feature extends to SGZ $> 40$ Mpc and expansion of the wall at our location is a significant source of the peculiar velocity of our Galaxy.  Indeed the Local Void is a prominent feature of the WF reconstruction.  The motion identified as `Local' in Table~\ref{table-vbulk} is in excellent agreement in both amplitude and direction with the local motion identified by Tully et al. (a motion of the Local Sheet of $323 \pm 25$ km/s toward $l_{gal} = 220 \pm 7$, $b_{gal} = +32 \pm 6$ or $sgl = 80$, $sgb = -52$).  The dominance of the Local Void marginalizes the dynamical role of the Virgo cluster on the local scene. Indeed, the structure within a Virgocentric sphere of a radius of $R=10.6 \hmpc$ contributes a mere 60 km/s which is directed $65^\circ$ away from the direction of the Virgo cluster. It follows that the Virgo cluster does not dominate even that Virgocentric sphere. 
Technically speaking, the current analysis is done within the linear regime and the results can be strictly implied only with that framework. However, the dynamical role of a virialized cluster at a distance of $\gtrsim 10 \hmpc$ away is only modestly affected by non-linear effects.

The motion of the LG with respect to the CMB is recovered by the present reconstruction to within 5\% in amplitude and 5 degrees in direction. 
Decomposing the velocity field into local and tidal components, with respect to the data zone sphere of $R=3000 \hmpc$, we find that the local structure contributes a component of  304 km/s  leaving a residual tidal component of  382 km/s, both deviating from the CMB dipole by $30^\circ$ but  in almost orthogonal directions. 

Next, the bulk velocity of the full data-zone has been examined. The bulk velocity of the full velocity field is very robustly determined to be $V_{bulk}=401 \pm 19$ km/s in the direction of  $l_{gal}=299\pm 6^\circ$ and $b_{gal}=24\pm6^\circ$. The statistical scatter is calculated over an ensemble of CRs. In principle, the bulk velocity can be partially induced by internal sources although the expectation is that it should be dominated by the exterior mass distribution. Indeed, the bulk velocity of the tidal field is found to be 378 km/s in the direction of $l_{gal}=293^\circ$ and $b_{gal}=18^\circ$. The shear tensor component of the velocity field does not have the characteristics that would be induced by a single attractor and none of the eigenvectors are closely aligned with the bulk velocity. 
It should be noted that the bulk velocity is defined here as the volume averaged velocity field within the region under consideration. This differs from the commonly defined bulk velocity, which is taken to be some weighted mean velocity of the galaxies that constitute a peculiar velocities data set. It follows that a direct comparison with the bulk velocity determinations of other surveys and by different methods should be done with care. Having stating this caveat we compare our estimated bulk velocity of the $R=30 \hmpc$ sphere with \citet{2011arXiv1101.1650N} who studied the SFI++ catalog and estimated the bulk velocity of a sphere of $R=40 \hmpc$ to be $333\pm38$ km/s toward $l_{gal}=276\pm3^\circ$ and $b_{gal}=14\pm3^\circ$. 
Given the very different methods of estimation and the different depth of the studies the two determination are in a good qualitative agreement. 

The WF/CR  algorithm presented here provides a robust methodology for recovering the underlying density and 3D velocity fields from observational surveys of the peculiar velocities of galaxies. Further progress in this field is to be pursued along two parallel tracks. On the one hand the method will be applied to a much larger and deeper sample of peculiar velocities measured in our `Cosmic Flows' program  \citep{2011MNRAS.tmp..466C}. This effort is to be matched by further theoretical work on the improvement of the method. The main shortcoming of the WF/CR method is the fact that it is strictly formalized within the linear regime. Our work-in-progress focuses on two main issues. One is the quest for a good grouping  algorithm of the data as a mean for suppressing virial velocities in collapsed objects and thereby 'linearizing' the data. The other is a dynamical mapping that enables the reconstruction of the quasi-linear density field from the WF linearly estimated field.

\section*{Acknowledgments}   
Comments by the referee have resulted in a significant improvement of the discussion.
HC has benefitted from the opportunity to spend a year at the Institute for Astronomy, University of Hawaii.
YH has been partially supported by the Israel Science Foundation (13/08).
RBT acknowledges support from the US National Science Foundation award AST09-08846. 
SG is partially supported by the German Academic Exchange Service (DAAD).


\bibliographystyle{mn2e}

\bibliography{HC}

\label{lastpage}

\end{document}